\documentclass{ws-p8-50x6-00}
\newcommand{\dd}{\displaystyle}

\begin{document}

\title{Effective Fields in Dense Quantum Chromodynamics}

\author{G. Nardulli}

\address{Dipartimento di Fisica, Universit\`a di Bari, I-70124
Bari, Italia \\ I.N.F.N., Sezione di Bari, I-70124 Bari, Italia\\
E-mail: giuseppe.nardulli@ba.infn.it}


\maketitle

\abstracts{In the high density, low temperature limit, Quantum
Chromodynamics
 exhibits a transition to phases characterized by color
superconductivity and energy gaps in the fermion spectra.  We
review some fundamental results obtained in this area and in
particular we describe  the low energy effective lagrangian
describing the motion of the quasi-particles in the high density
medium (High Density Effective Theory).}

\section{Lecture I: Color superconductivity}
\subsection{Introduction}\label{subsec:prod}At high baryonic
densities and small temperatures
 the color interaction favors the formation of
quark-quark condensates in the color attractive antisymmetric
channel: \be\Delta\,=\,\langle\psi^T_{i\alpha }C\psi_{j\beta}
\rangle\,\epsilon^{\alpha\beta\gamma}\epsilon^{ijk}\Omega_{\gamma
k }\,\neq\,0
 \label{condensatedue}\ee ($\alpha,\,
\beta,\,\gamma=1,2,3$ color indices; $i,\,j,\,k=1,2,3$ flavor
indices). The condensates (\ref{condensatedue}) depend on the
matrix   $\Omega$ and act as order parameters of new phases where
the $SU(3)_c$ color symmetry is spontaneously broken. The
densities are so high that these phenomena might probably occur
only in the core of neutron stars. Since this mechanism is similar
to electron superconductivity it is referred to as color
 superconductivity (CSC) \cite{others}. This  is one of the most fascinating
 advances in Quantum Chromo Dynamics (QCD) in recent years
 \cite{alford,wilczekcfl,wilczekcfl1}
 %,wilczekcfl2,wilczekcfl3,wilczekcfl4,Schafer:1998na,generali,shuryak,alford1,pisarski,altri,kogut}
 (for reviews see
 \cite{rassegne}).
 %,rassegne2,hsu}).
 The aim of these lectures  is to describe an  approach  to this aspect of
 QCD  that is  based on the method of the effective  lagrangians
  \cite{cflgatto,2fla,gattoloff,nardullinc}
 and to stress  possible astrophysical consequences.

 Different phenomena take place depending on the value of
the order parameter (\ref{condensatedue}). One could have: $
\Omega_{\gamma k}=\delta_{\gamma 3}\delta_{k 3}$, which
corresponds to
\be\epsilon^{\alpha\beta\gamma}\epsilon_{ij}<\psi^T_{i\alpha
}C\psi_{j\beta}>=\Delta\delta^{\gamma 3}\label{lab2}\ee where the
sum over the flavor indices run from 1 to 2 and $\psi$ represents
a left handed 2-component Weyl spinor (the right handed field
satisfies a similar relation); moreover a sum over spinor indices
is understood and $C=i\sigma_2$. This case correspond to
decoupling of the strange quark ($m_s\to\infty;\, m_u=m_d=0$) and
is called 2SC model. From dynamical analyses
\cite{alford,rassegne} one knows that, for $\mu$ sufficiently
large, the condensate (\ref{lab2}) is non vanishing. Therefore it
breaks the original symmetry group $SU(3)_c\otimes SU(2)_L\otimes
SU(2)_R\otimes U(1)_B$ down to $SU(2)_c\otimes SU(2)_L\otimes
SU(2)_R\otimes Z_2$.  The chiral group remains unbroken, while the
original color symmetry group is broken to $SU(2)_c$, with
generators $T^A$ corresponding to the generators $T^1,T^2,T^3$ of
$SU(3)_c$; an unbroken $U(1)_{\tilde B}$ also remains. As a
consequence, three gluons remain massless whereas the remaining
five acquire a mass. One can construct an effective theory to
describe the emergence of the unbroken subgroup $SU(2)_c$ and the
low energy excitations, much in the same way as one builds up
chiral effective lagrangian with effective fields at zero density.
For the two flavor case this development can be found in
 \cite{casalbuonisannino}.

For the three flavor case ($m_u=m_d=m_s=0$)  the following case
has been widely discussed \cite{wilczekcfl}: \be
\langle\psi_{i\alpha}^L\psi_{j\beta}^L\rangle=
-\langle\psi_{i\alpha}^R\psi_{j\beta}^R\rangle=\Delta
\sum_{K=1}^3\epsilon_{\alpha\beta K}\epsilon_{ijK}~
.\label{condensates}\ee The condensate (\ref{condensates}) breaks
the original symmetry group $SU(3)_c\otimes SU(3)_L\otimes
SU(3)_R\otimes U(1)_B$ down to $SU(3)_{c+L+R}\otimes Z_2$. Both
the chiral group, $U(1)_B$ and the  color symmetry are broken but
a diagonal $SU(3)$ subgroup remains unbroken in a way that locks
together color and flavor (Color-Flavor-Locking=CFL model). There
are 17 broken generators; since there is a broken gauge group,
 8
of these generators correspond to 8 longitudinal degrees of
freedom of the gluons, because the gauge bosons acquire a mass;
there are 9 Nambu-Godstone Bosons (NGB's) organized in an octet
associated to the breaking of the flavor group and in a singlet
associated
 to the breaking of the baryonic number.
 The effective theory describing the NGB for the CFL model
 was studied in \cite{casalbuonigatto}.

Another interesting possibility  arises when there is a difference
$\delta\mu$ between the chemical potentials of the two gapped
quarks (for references  see \cite{LOFF}; see also
\cite{bowers,crystal} and for earlier works \cite{originalloff}).
As shown in \cite{LOFF} when the two fermions have different
chemical potentials $\mu_1\neq \mu_2$, for $\delta\mu$ of the
order of the gap the vacuum state is characterized by a non
vanishing expectation value of a quark bilinear which breaks
translational and rotational invariance. The appearance of this
condensate is a consequence of the fact that for $\mu_1\neq\mu_2$,
and in a given range of $\delta\mu=|\mu_1-\mu_2|$, the formation
of a Cooper pair with a total momentum $\vec p_1\,+\,\vec
p_2\,=\,2\vec q\neq \vec 0$
 is energetically favored in comparison with the normal
BCS state. The possible form of these condensates is discussed in
\cite{LOFF} and \cite{bowers} (see also \cite{crystal}); it is
worthwhile to  note that, for simplicity, these authors assume
only two flavors. In \cite{LOFF} the ansatz of a plane wave
behavior for the condensate is made : $\dd\propto e^{2i\vec
q\cdot\vec x}$; in \cite{bowers} it is shown that the
configuration that is energetically favored is a face centered
cubic crystal. For simplicity I will consider here only the plane
wave:
 \be
-<0|\epsilon_{ij}\epsilon_{\alpha\beta 3 } \psi^{i\alpha}( \vec
x)C\psi^{j\beta}(\vec x)|0>= 2\Gamma_A^L e^{2i\vec q\cdot\vec
x}\label{scalar}\ ;\ee besides the scalar condensate
(\ref{scalar}) there is also a  vector condensate that can be
however neglected since it is numerically small. The 2SC and/or
the LOFF phases  might exist  for intermediate values of the
chemical potentials, while for very high $\mu$ the CFL phase
should set in.

The phase transition and the non vanishing  condensates
 result from a mechanism
analogous to the formation of an electron Cooper pair in a BCS
superconductor. At $T=0$ the only QCD interactions are those
involving fermions near the Fermi surface. Quarks inside the Fermi
sphere cannot interact because of the Pauli principle, unless the
interactions involve large momentum exchanges. In this way the
quarks can escape the Fermi surface, but these processes are
disfavored, as large momentum transfers imply small couplings due
to the asymptotic freedom property of QCD. Even though
interactions of fermions near the Fermi surface involve momenta of
the order of $ \mu$, their effects are not necessarily negligible.
As a matter of fact, even a small attractive interaction between
fermions near the Fermi surface and carrying opposite momenta can
create an instability and give rise to coherent effects. This is
what really happens \cite{alford,wilczekcfl} and the result is the
formation of a diquark condensate, as expressed by (\ref{lab2}),
(\ref{condensates}) or (\ref{scalar}). We stress again that the
only relevant fermion degrees of freedom are therefore those near
the Fermi surface. In \cite{cflgatto} an effective theory for the
CFL model was discussed, based on the approximation of the neglect
of the negative energy states. This results in a rather terse
formalism displaying as a characteristic note the existence of a
Fermi velocity superselection rule and effective
velocity-dependent fermion fields. We will refer to this effective
lagrangian as the High Density Effective Theory (HDET). In
\cite{2fla} the 2SC model has been studied by the same formalism,
while in \cite{gattoloff}
%,nardulliloff,gattoloff2}
this effective theory has been applied to  the crystalline color
superconducting phase \cite{LOFF}, the so-called LOFF
\cite{originalloff} phase.

 The aim of
these lectures  is to review some  developments in the description
of Color Super Conductivity that are based on the  HDET approach.
This will be mainly done in the second lecture. Here I wish to
discuss some possible astrophysical implications of CSC, in
particular in the LOFF phase.\subsection{Astrophysical
implications of the LOFF phase}

Besides its theoretical interest for the study of the phase
structure of QCD theory, the crystalline phase may result relevant
for astrophysical dense systems, in particular in the explanation
of the glitches in the pulsars.

The pulsars are rapidly rotating stellar objects, characterized by
the presence of strong magnetic fields and by an almost continuous
conversion of rotational energy into electromagnetic radiation.
The rotation periods can vary in the range $10^{-3}$ sec up to a
few seconds; these periods increase slowly and  never decrease
except for occasional glitches, when the pulsar spins up with a
variation in frequency that can be   $ \delta\Omega/\Omega\approx
10^{-6}$ or smaller. Glitches are a typical phenomenon of the
pulsars, in the sense that probably all the pulsar have glitches.

Pulsar are commonly identified with neutron stars; these compact
stars are characterized by a rather complex structure comprising a
core, an intermediate region with superfluid neutrons and a
metallic crust. The ordinary explanation of the glitches is based
on the idea that these sudden jumps of the rotational frequency
are due to the movements outwards of rotational vortices in the
neutron superfluid and their interaction with the crust. This is
one of the main reasons that allow the identification of pulsars
with neutron stars, as only neutron stars are supposed to have a
metallic crust. Since the conventional models for glitches  may be
not familiar to an audience of nuclear and/or high energy
physicists, I will briefly review them in the sequel.

It is known that a boson liquid at $T\sim 0$ forms a condensate
whose wavefunction \be\Xi_0=\sqrt{n_0(t,\vec r)}\,e^{i\Phi(t,\vec
r)}\ee has  a macroscopic meaning, due to the large number of
particles in it. Also the probability current density
\be\displaystyle\vec j_{cond}=\frac{i\hbar}{2m}
(\Xi_0\nabla\Xi_0^*-\Xi_0^*\nabla\Xi_0)=
\frac{\hbar}{m}n_0\nabla\Phi\ee  has such macroscopic meaning.
Since $ \vec j_{cond}=n_0\vec v_s$, where $\vec v_s$ is the
condensate velocity that can be identified with the superfluid
velocity, we get \be \vec v_s=\frac{\hbar}{m}\,\vec \nabla\Phi\
.\label{vs}\ee The consequence of (\ref{vs}) is
\be\oint_\gamma\vec v_s\cdot d \vec\ell=0\ ,\label{g50}\ee if the
domain where the curve $\gamma$ lies is simply connected. Given
the arbitrariness of $\gamma$, a  different way to write this
result is  \be \vec\nabla\wedge\vec v_s=0\label{g50bis}\ .\ee

Let us now suppose that the vessel containing the liquid is put in
rotation with angular velocity $\vec\Omega$. A consequence of
(\ref{g50}) is that the superfluid cannot rotate; to the same
result one arrives by noting that the vessel cannot communicate
the rotation to the superfluid component, as there is no friction
between the recipient and the liquid. However it can be shown that
the absence of rotation in the superfluid does not correspond to a
state of minimal energy. As a matter of fact, if $E$ and $\vec L$
are energy and angular momentum in an inertial frame, the energy
as computed in the rotating frame is $E_{rot}=E-\vec L\cdot\vec
\Omega$. If $\Omega$ is sufficiently high, then a lower energy can
be achieved with \be\vec L\cdot\vec \Omega>0 \label{om}\ee instead
of $\vec L\cdot\vec \Omega=0$ that corresponds to the absence of
rotation \cite{LL}.

Let us suppose for simplicity that the curve $\gamma$ lies in a
plane. For the result (\ref{om}) to be compatible with (\ref{g50})
one has to suppose that inside the curve $\gamma$ there is one
point where (\ref{g50bis}) is violated. Physically this would
correspond to the presence of a point with a normal, not
superfluid component. Since now the domain is not simply
connected, (\ref{g50}) is substituted by \be\oint_\gamma\vec
v_s\cdot d \vec\ell\,=\,2\pi n\kappa\ ,\label{g5}\ee where the
integer $n$ is a winding number counting the number of times the
curve goes around the singular point, and $\kappa$ is a constant
with dimensions of {\it vorticity}, i.e. $[L]^2\cdot[T]^{-1}$. To
the same result one could arrive by noting that, by virtue of
(\ref{vs}), as $\Phi$ and $\Phi+2n\pi$ correspond to the same
wavefunction, one may have \be\oint_\gamma\vec \nabla\Phi\cdot d
\vec\ell\,=\,2\,\pi \,n\, ,\label{g6}\ee which shows that $
\kappa={\hbar}/{m}$; $\kappa$ is called quantum of vorticity.

Clearly we can repeat the argument for any plane parallel to the
previous one; we therefore conclude that there is an entire line
({\it vortex line}) of singular points. If this line is a straight
line, then $\vec v_s$ will be perpendicular to the vortex line and
also perpendicular to the radius joining the singular point and
the point at which we compute $v_s$. At a distance $r$ from the
singular point one has\be v_s=\frac{n \kappa}{r}\ ,\label{g7}\ee
as can be immediately seen from (\ref{g5}). More generally: \be
\vec v_s=\frac{\kappa}{2}\int_{v.l.}\frac{d\vec\ell\wedge\vec
R}{R^3}\ ,\ee where $\vec R$ is the distance vector from the
vortex line (v.l.) to the point at which we compute the superfluid
velocity.

We note some properties of the vortex lines. First, the integral
(\ref{g5}) is independent of $\gamma$, provided the second curve
contains the singular point; this result follows from the Stokes
theorem. Second, the vortex line must be closed or it must stop at
the boundary; were it open, for example at a point P, one might
construct a surface $\Sigma$ lying on the contour $\gamma$ but
large enough as to have no intersection with the vortex line;
therefore  by the Stokes theorem one would get that the vorticity
constructed by   $\vec v_s$ would be zero. If the vortex line
stops at the boundary and the vessel is rigid then the v.l. is
pinned at the boundary, as it will be discussed in more detail
below.

 Let us compute  the
critical angular velocity $\Omega$ for the formation of the first
vortex line. The formation of a vortex line changes the energy
$E_{rot}$ by the amount $ \Delta E_{rot}=\Delta E-\Delta
\left(\vec L\cdot \vec\Omega\right)$. We have $\displaystyle\Delta
E=\int\frac{\rho_sv_s^2}{2}dV$. Here the integral is over the
entire volume of the vessel that, for simplicity, we assume to be
a cylinder of height $b$ and radius $R$; on the other hand $v_s$
is given by the expression valid for a vortex line, i.e.
(\ref{g7}). Therefore \be \Delta E=\frac{\rho_sb}{2}\,2\pi\,\int
v_s^2 r dr =b\rho_s\pi\frac{\hbar^2n^2}{m^2}\ln\frac{R}{a}\ee
where $a$ is a cutoff of the order of the interatomic distances,
at which the macroscopic description fails down. The minimal
energy $E_{rot}$ is obtained for $\vec L$ parallel to $\vec
\Omega$, with $\displaystyle L= \int\rho_sv_s r\,dV=b\pi
R^2\frac{n\hbar}m\rho_s$. Vortex lines appear if $\Delta
E_{rot}<0$, i.e. if\be
 \frac{n\hbar}m\ln\frac R
a-\Omega\,R^2<0\label{g8}, \ee which corresponds to $\displaystyle
\Omega>\Omega_{crit}=\frac\hbar {mR^2}\ln\frac R a$. Incidentally
we note that, {\it ceteris paribus}, vortex lines with $n=1$ are
more stable than those with $n>1$, as the positive term in
(\ref{g8}) has an extra power of $n$. Therefore from now on we put
$n=1$.

What happens when $\Omega\gg\Omega_{crit}$? Clearly we expect
 there will be several v.l.'s. During the rotation these vortex
lines follow the rotational motion of the vessel, which is clear
because they are pinned at the boundary of the
superfluid\footnote{For rotations around an axis, the vortex lines
are, by symmetry, straight lines parallel to the rotation axis.
For motion inside holes, slits , etc. there can be closed v.l.'s
that are called {\it vortex rings.}}. Their motion imitate the
motion of the liquid as a whole, as it can be seen by the
following argument: in the formula for $\Delta E_{crit}$ one can
forget, for large $\Omega$, the first term $\Delta E$ in
comparison to the second one. Therefore minimizing $\Delta
E_{crit}$ correspond to maximize the angular momentum $L$, which
is obtained if the liquid moves as a whole. A consequence of this
feature of the motion of the v.l.'s is that also for the
superfluid one can use the well known hydrodynamical formula
$\displaystyle \vec\Omega=\frac 1 2\vec\nabla\wedge\vec v_s$,
which strictly speaking is valid only for the normal component
$\vec v_n$; it can be used here only because it refers to the
rotation of the superfluid as a whole.

The number of vortex lines that are present in the superfluid is
proportional to $\Omega$, according to the formula ($N$=number of
lines  per unit area):\be N=\frac{m\Omega}{\pi\hbar} \label{gn}\ee
which shows that with increasing $N$ the v.l.'s tend to fill in
all the space. To prove (\ref{gn}) we consider a large closed
curve $C$ encircling the area $A$ and containing in its interior
$NA$ v.l.'s. One has \be\displaystyle \oint_Cd\vec\ell\cdot\vec
v_s=NA2\pi\kappa\label{n}\ee but also \be\oint_Cd\vec\ell\cdot\vec
v_s= \int_A d\vec\sigma\cdot \vec\nabla\wedge\vec v_s=2 A\Omega\ee
from which (\ref{gn}) follows. As an example we can evaluate $N$
for the pulsar in the Crab nebula. Here $m=2m_N$ (the condensate
is formed by neutral bosons: pairs of neutrons) and
$\Omega=\Omega_{pulsar}$ gives $ N\simeq 1.9\times 10^{5}cm^{-2}$
with an average distance between vortex lines $d\sim N^{-1/2}\sim
10^{-2}cm$.

 Let us consider
again eqns. (\ref{g6}) and (\ref{n}). If $\nu(r) $ is the number
of vortices per unit area at a distance $r$ from the rotation
axis, they give, if $\vec v=\vec v_s$ is the superfluid velocity,
 \be
\oint d\vec\ell\cdot\vec v=\int_0^r d\vec
S\cdot\vec\nabla\wedge\vec v=2\pi\kappa\int_0^r 2\pi
r^\prime\nu(r^\prime)dr^\prime\ .\label{alp1}
 \ee
 We put
 \be
 k=2\pi\kappa=\frac h{2m_n}\ee
 and write (\ref{alp1}) as follows:
 \be
2\pi\,r^2\,\Omega(r)=k\,\int_0^r 2 \pi
r^\prime\nu(r^\prime)dr^\prime\ ,\label{alp2}
 \ee
 which implies
 \be
k\nu(r)=2\Omega(r)\,+\,r\,\frac{\partial\Omega}{\partial r}\ .
 \label{alp4}\ee
 Since the  total number
 of v.l.'s is conserved, one has
 \be\frac{\partial \nu}{\partial t}+\vec\nabla\cdot(\nu\vec
 v_r)=0\ee
where $\vec v_r$ is the radial component of the superfluid
velocity. We write (\ref{alp2}) as\be
2\pi\,r^2\,\Omega(r)=k\,\int_0^r\, \nu\,dS \label{alp3}
 \ee and take the time derivative, using
(\ref{alp4}) to get\be 2\pi\,r^2\,\frac{\partial\Omega}{\partial t
}=-k\,\int_0^r dS \vec\nabla(\nu\vec v_r)\ .\label{alp5}
 \ee Using the Gauss theorem one gets
 $\displaystyle 2\pi\,r^2\,\frac{\partial\Omega}{\partial t
}=-k\,2\pi r \nu v_r$,
 i.e.
\be \frac{\partial\Omega}{\partial t }=-\frac{k \nu
v_r}r=-\left(2\Omega(r)\,+\,r\,\frac{\partial\Omega}{\partial
r}\right)\,\frac{v_r}r \ .\label{alp7}
 \ee
 Eq. (\ref{alp7}) shows that {\it the only possibility for the
 superfluid to change its angular velocity} ($\dot\Omega\neq 0$) {\it
 is by means of a
 radial motion, i.e. $v_r\neq 0$}.

 Let us now consider a rotating superfluid in contact with rotating normal matter
 on which an external torque is acting \cite{alpar}. We denote by
 $I_c,\,\Omega_c$ moment of inertia and angular velocity of the
 normal components  that, in a neutron star, includes
  the crust and possibly other normal components.
  The equation of motion of the normal component is
\be I_c\dot\Omega_c(t)=M_{ext}+M_{int}\ .\label{alp8}\ee Besides
the external torque $M_{ext}$, basically related to the spin down
of the pulsar
 (or the steady accretion in binary pulsars), we have included internal torque
 $M_{int}$:
 \be
 M_{int}=-\int dI_p\,\dot\Omega (r,t)\label{alp9}
 \ee due to the interaction with the superfluid.
 Eqs.(\ref{alp7}-\ref{alp9}) are the equations of
 motion for the
 angular velocities $\Omega$ and $\Omega_c$ (superfluid and crust).
  The two velocities  are coupled not
 only through $M_{int}$, but also by $v_r$, because we will show
 below that $v_r=f(\Omega-\Omega_c)$. We note again that
 fundamental for this model is the existence of radial motion,
 for, if $v_r=0$, then $\Omega= const.$ and only $\Omega_c$
 changes, due to the external torque alone.

 In the neutron star, superfluid neutrons (in Cooper pairs)
 coexist with nuclei of the crust. Also in the crust there are
superfluid neutrons, but they are characterized by a different
(and smaller)
 $\Delta$. For superfluid neutrons in the volume $V$, the
 total energy can be
 estimated as follows:
 \be
 E\simeq \frac{V}{(2\pi\hbar)^3}\int_{p_F}^{p_F+\Delta^2/E_F}p^3dp
 \frac{d\Omega}{2}=
 \frac{V}{4\pi^2}\frac{\Delta^2k_F^3}{E_F}\ee
 where
 $p_F=\hbar k_F$; we integrate over half the solid angle
 as the superfluid  neutrons only appear in pairs; we have taken into account
 that only neutrons
 in a shell of thickness $\Delta^2/E_F$ participate in the pairing.

$E$ is also approximately given by the difference between the
energies of superfluid neutrons outside the vortex line and
neutrons inside, because those inside the vortex core have
$\Delta\to 0$. Therefore neutrons inside the volume $V$ of the
vortex core are repelled from going outside the vortex towards the
superfluid phase as it would cost more energy. However, if neutron
rich nuclei are present, the repulsion will be less important, as
$\Delta_c$, the gap for superfluid neutrons in the nuclei, is much
smaller than $\Delta_s$, the gap of superfluid neutrons; therefore
there will be a force pulling the vortex toward the nuclei; the
pinning energy per nucleus will be
 \be\delta E_p=
 \frac{V}{8}\left[\left(\frac{\Delta^2k_F^3}{\pi^2E_F}\right)_s
 -\left(\frac{\Delta^2k_F^3}{\pi^2E_F}
 \right)_c\right]\approx\gamma \frac{V}{8}
 \left(\frac{\Delta^2k_F^3}{\pi^2E_F}\right)_s
 \ee with $\gamma\sim 1$.

 Let now $\xi$ be the {\it coherence length}, i.e. the spatial
 extension of the Cooper pair,
$\displaystyle \xi=
 \frac{\hbar v_F}{\pi\Delta}$; it can be proved that it gives an
 estimate of the radius of the vortex core. The maximum pinning
 force will be obtained, if $2\xi<b$ ($b$ the average distance
 between the nuclei) when the vortex passes through one layer of
 the lattice; the average distance between vortex core neutrons and superfluid neutrons
 is of the order of $\xi$ and therefore the maximum force is
 $\displaystyle F_p\simeq \frac{\delta E_p}{\xi}$ and
 the maximum force per unit lenght of vortex line ($b$ is also the average
 distance between two consecutive pinning centers)   is
 \be
 f_p\simeq \frac{\delta E_p}{b\xi}\ .\ee

 Let us finally discuss a possible mechanism for the formation
  of glitches \cite{alpar}; we
  consider the rotating neutron star with superfluid
 neutrons in its interior and a metallic crust, which is a
 simplified model, but adequate to our purposes. We
 distinguish between two angular velocities: the superfluid
 velocity  $\Omega$ and the crust velocity $\Omega_c$. Let us
 suppose that they are initially equal, which is a consequence of
 the pinning.
  Due to the spinning down
 of the star, $\Omega_c$ decreases; as long the vortex cores are
 pinned to the crust lattice, the neutron superfluid cannot spin down,
  because the radial motion is forbidden. There is therefore a
  relative velocity of the superfluid with respect
  to the  pinned vortex core because $\Omega>\Omega_c$:
  \be \delta\vec v=(\vec \Omega-\vec\Omega_c)\wedge \vec r\ .\ee The interaction
  between the normal matter in the core of
  the v.l. and the rest of normal matter (nuclei in the lattice,
  electrons, etc.) produces a Magnus force per unit length
  given by
 \be\vec f=\rho \vec k\wedge\delta\vec v\ ,\ee
where $k$  is the quantum of vorticity and the direction of $\vec
k$ coincides with the rotation axis. $f$ is the force exerted on
the vortex line; as it cannot be larger than $f_p$ there is a
maximum difference of angular velocity that the system can
maintain: \be \omega_{cr}=\left(\Omega-\Omega_c\right)_{max}=
\frac{f_p}{\rho k r}=\frac{E_p}{\rho k\xi b}\ .\ee If $\omega
<\omega_{cr}$ the vortices remain pinned at the lattice sites
instead of flowing with the superfluid as they generally in
superfluid (see discussion above). On the contrary, if $\omega
>\omega_{cr}$, the hydrodynamical forces arising from the mismatch
between the two angular velocities ultimately break the crust and
produce the conditions for the glitch.

We stop here this introduction to the standard model for glitches;
see \cite{alpar} for more details. The relevance for CSC is that
the LOFF phase  provides a lattice structure independently of the
crust. Even in quark stars, if one is in a LOFF phase, one has a
crystal structure: a lattice characterized by a geometric array
where the gap parameter varies periodically. This would avoid the
frequently raised objection by which one excludes the existence of
strange stars because, if the strange matter exists, quark stars
should be rather common, in contrast with the widespread
appearance of glitches in pulsars. Therefore, if the color
crystalline structure is able to produce glitches, the argument in
favor of the existence of strange stars would be reinforced.

 In a more conservative vein one can
also imagine that the LOFF phase be realized in the inner core of
a neutron star; in this case the crystalline color
superconductivity could be partly responsible for the glitches of
the pulsar. A detailed analysis of this scenario is however
premature as one should first complete the study of the LOFF phase
by including the third quark and by sorting out the exact form of
the color lattice \cite{bowers,crystal}.

\section{Lecture II: High Density Effective Theory }
In this section we derive the general formalism of High Density
Effective Theory and we present an example of its use. This
effective theory shows some resemblance with the Heavy Quark
Effective Theory (see e.g. \cite{HQET}); it is discussed in detail
in the review paper \cite{nardullinc}.
\subsection{General Formalism \label{ch4.1}}
  The main idea of the effective theory is the
observation that the quarks participating in the dynamics have
large ($\sim \mu$) momenta. Wherefore one can introduce velocity
dependent fields by extracting the large part $\mu\vec v$ of this
momentum. One starts with the Fourier decomposition of the quark
field $\psi(x)$: \be
\psi(x)=\int\frac{d^4p}{(2\pi)^4}e^{-i\,p\cdot x}\psi(p)\
,\label{4.0.dec}\ee and introduces the quark velocity by
 \be p^\mu=\mu
v^\mu+\ell^\mu\ , \label{4.1.dec} \ee where $v^\mu=(0,\vec v )$
with $|\vec v|=1$. Let us put $ \ell^\mu=(\ell^0,\vec\ell) $ and
$\vec\ell=\vec v \ell_\parallel+\vec\ell_\perp$ with
$\vec\ell_\perp= \vec\ell-(\vec\ell\cdot\vec v )\vec v$. We can
always choose the velocity parallel to $\vec p$, so that
$\vec\ell_\perp=0$ and \be \int d^4 p=\mu^2\int d\Omega\int
d\ell_\parallel\int_{-\infty}^{+\infty}d\ell_0=
4\pi\mu^2\int\frac{d\vec v}{4\pi} \int  d\ell_\parallel
\int_{-\infty}^{+\infty} d\ell_0 \ .\label{4.0.8}\ee
  In this way
   the Fourier decomposition (\ref{4.0.dec}) takes the form
\be \psi(x) = \sum_{\vec v} e^{-i\mu v\cdot
x}\int\frac{d^4\ell}{(2\pi)^4}e^{-i\ell\cdot x}\,\psi_{\vec v
}(\ell)= \sum_{\vec v} e^{-i\mu v\cdot x}
\frac{4\pi\mu^2}{(2\pi)^4} \int d^2\ell\,e^{-i\ell\cdot
x}\,\psi_{\vec v }(\ell) ,\ee
%\ ,\ee where\be
 where $\psi_{\vec v }(\ell)$ are velocity-dependent
fields.  One can write\be\psi(x)=\sum_{\vec v} e^{-i\mu v\cdot x}
 \left[\psi_+(x)+\psi_-(x)\right]\ ,\ee where
 $\psi_\pm$ are velocity dependent fields
corresponding to positive and energy solutions of the Dirac
equation.

Let us now define $V^\mu=(1,\,\vec v)$, $\tilde V^\mu=(1,\,-\vec
v)$, $\gamma^\mu_\parallel=(\gamma^0,\,(\vec v\cdot\vec
\gamma)\,\vec v)$,
$\gamma_\perp^\mu=\gamma^\mu-\gamma_\parallel^\mu$. Using simple
algebraic relations involving the gamma matrices \cite{nardullinc}
one obtains \be {\cal L}_D=\sum_{\vec v}{} \Big[\psi_+^\dagger
iV\cdot D\psi_++\psi_-^\dagger(2\mu+ i\tilde V\cdot D)\psi_-\,+
(\bar\psi_+iD_\mu\gamma^\mu_\perp\psi_- + {\rm h.c.} )\Big]\ ;\ee
$D_\mu$ is the covariant derivative: $D^\mu=\partial^\mu+ig
A^\mu$. We note that here quark fields
 are   evaluated at the same Fermi velocity; off-diagonal terms
are subleading due to the Riemann-Lebesgue lemma, as they
 are cancelled by the rapid oscillations of the
exponential factor in the $\mu\to\infty$ limit. One may call this
phenomenon {\it  Fermi velocity superselection rule}, in analogy
with the behaviour of QCD in the $m_Q\to\infty$ limit, where the
corresponding effective theory, the Heavy Quark Effective Theory
exhibits a similar phenomenon \cite{HQET}. By the same analogy we
may refer to the present effective theory as High Density
Effective Theory (HDET),

We can get rid of the negative energy solutions by integrating out
the $\psi_-$ fields in the generating functional; in this way we
get\be {\cal L}_D\simeq{\cal L}_{0}= \sum_{\vec v}
\Big[\psi_+^\dagger iV\cdot
\partial\psi_+\ +\
\psi_-^\dagger i\tilde V \cdot \partial\psi_-\Big]\ ,\ee where now
$\psi_\pm$ are both positive energy solutions with $\psi_\pm
=\psi_{\pm\vec v}$
 The construction described above  is  valid for any theory
describing massless fermions at high density provided one excludes
 degrees of freedom very far from the Fermi surface. As discussed
 in the first lecture however, for small temperature and high  density
  the fermions are likely to be gapped due to
  the phenomenon of the color superconductivity. We shall examine
  here the
  modification in the formalism
for the LOFF model, with the condensate
 \be \Delta(\vec x)=\Delta\,\exp\{2i\vec
q\cdot\vec x\}\ . \label{1}\ee
 The effect of the non vanishing vacuum
expectation value can be taken into account by adding to the
lagrangian the term: \be {\cal
L}_\Delta=-\frac{\Delta}2\,\exp\{2i\vec q\cdot\vec
x\}\,\epsilon_{\alpha\beta 3} \epsilon_{ij}\psi_{i\alpha}^T(x)C
\psi_{i\beta}(x)\,-(L\to R)+{\rm h.c.}\ .\label{ldeltaeff}\ee

 In order to introduce velocity dependent positive energy
fields $\psi_{\vec v_i;\,i\alpha}$ with flavor $i$ we decompose
both fermion momenta as in (\ref{4.1.dec}) and we have: \bea {\cal
L }_\Delta&=&-\frac{\Delta} 2\, \,\sum_{\vec v_i,\vec v_j}
\exp\{i\vec x\cdot\vec\alpha(\vec v_i,\,\vec v_j,\,\vec q
)\}\epsilon_{ij}\epsilon_{\alpha\beta 3}\psi_{-\,\vec
v_i;\,i\alpha}(x)C \psi_{-\,\vec v_j;\,j\beta}(x)\cr && -(L\to
R)+{\rm h.c.}\ ,\label{loff6}\eea where $\vec\alpha(\vec
v_i,\,\vec v_j,\,\vec q)=2\vec q-\mu_i\vec v_i-\mu_j\vec v_j$. We
define
\be\mu=\frac{\mu_1+\mu_2}{2}~,~~~~~~~~~\delta\mu=\,-\,\frac{\mu_1-\mu_2}{2}
,\label{dec2}\ee and perform the $\mu\to\infty$ limit on a smeared
amplitude as follows \be\lim_{\mu\to\infty}\exp\{i\vec
x\cdot\vec\alpha(\vec v_1,\,\vec v_2,\,\vec q)\} \equiv \frac 1 V
\int_{V(\vec x)}d\vec r\, \exp\{i\vec r\cdot\vec\alpha(\vec
v_1,\,\vec v_2,\,\vec q)\} ,\label{1bis}\ee where ${V(\vec x)}$ is
a small volume centered at $\vec x$. We evaluate (\ref{1bis}) by
taking $\vec q$ along the $z-axis$, so  that  we get: $ \vec
v_i\simeq -\vec v_j\equiv -\vec v$ by the $x$ and $y$
integrations, while the z-integration gives ($\vec n=\vec q/q$)
\be \frac{\pi}{R}e^{i2qhz}\,\delta_R[h(\vec v\cdot\vec
n)]\approx\frac{\pi}{R} \delta_R[h(\vec v\cdot\vec n)].\ee
  We have put  $ R=q|\Delta \ell|$
  where $\Delta\ell$ is a smearing distance
  along the direction of $\vec q$ ($\dd |\Delta \ell|
  \sim\frac{\pi}{q}$). We have
   introduced the "fat delta" $\delta_R(x)$  defined by
 \be
\delta_R(x)\equiv\frac{\sin(Rx)}{\pi x} \ ,\label{tre}\ee which,
for large $R$, gives $\delta_R(x)\to \delta(x)$. Moreover in the
$\mu\to\infty$ limit $\displaystyle h(x)=1-\frac{\delta\mu}{qx}$.
Therefore one has \be {\cal L }_\Delta=-\frac{\Delta} 2\, \frac\pi
R\,\delta_R[h(\vec v\cdot\vec n)]\sum_{\vec
v}\epsilon_{ij}\epsilon_{\alpha\beta 3}\psi_{\vec v;\,i\alpha}(x)C
\psi_{-\,\vec v;\,j\beta}(x)-(L\to R)+{\rm
h.c.}.\label{loff6bis}\ee
 In an appropriate basis
 the effective lagrangian is
 \be  {\mathcal L}= {\mathcal L}_0 \ +\ {\mathcal L}_\Delta
 = \sum_{\vec v}\sum_{A=0}^5 \chi^{A\,\dag}\left(
\begin{array}{cc}
 i\, \delta_{AB}\ V\cdot\partial\ & \,\Delta_{AB}
\\
\,  \Delta_{AB} & i\,\delta_{AB}\ \tilde V\cdot\partial\
\end{array}\right)\chi^B\ ,\ee
where the  matrix $\Delta_{AB}$ is as follows: $ \Delta_{AB}=0$
$(A\,{\rm or} \,B=4\,{\rm or} \,5 )$, and, for $A,B=0,...,3$:\be
\Delta_{AB}=\Delta_{eff} \left(\begin{array}{cccc}
   1& 0 & 0 & 0 \\
  0 & -1 &0 & 0 \\
  0 & 0&-1& 0 \\
  0 & 0 & 0 & -1
\end{array}\right)_{AB}\ ,\label{eq:38}
\ee having put\be\Delta_{eff}=\frac{\Delta\pi} R\delta_R[h(\vec
v\cdot\vec n)]\ . \label{deltaeff}\ee From these equations one can
derive the  propagator for gapped quarks:
 \be D_{AB}(\ell)=\frac{1}{V\cdot
\ell\,\tilde V\cdot \ell\,-\,\Delta_{eff}^2} \left(
\begin{array}{cc}
\tilde V\cdot\ell\,\delta_{AB} &  -\Delta_{AB}
\\
 -\Delta_{AB}
 &
 V\cdot\ell\,\delta_{AB}
\end{array}
\right)\label{propagatore}\ .\ee
\subsection{NGB
and their parameters}Both in the CFL model and in the LOFF model
with two flavors there are Nambu-Goldstone Bosons, associated in
one case to the breaking of internal symmetries and in the other
to the breaking of space-symmetries (phonon).
 In order to derive an effective low energy lagrangian for the
NGB's  one can use the gradient expansion, where
  the NGB's are introduced as external
fields and acquire a
 kinetic term, thus becoming dynamical fields, by integrating
 out the fermion fields.
I shall describe in some
 detail the calculation for the phonon field in the LOFF phase
 (other examples can be found in \cite{nardullinc}).

 ${\cal L}_\Delta$ in (\ref{ldeltaeff})
 explicitly breaks rotations and translations
and induces a  lattice structure given by plane waves. The crystal
can fluctuate and its local deformations define one phonon field
$\phi$ that is the Nambu-Goldstone boson associated to the
breaking of the translational symmetry. It is introduced by the
substitution in (\ref{1}) \be z\to
z+\frac\phi{2qf}\,,\label{small}\ee with $\langle\phi\rangle_0=0$.
We are interested in an effective description of the field $\phi$
in the low energy limit, i.e. for wavelengths much longer than the
lattice spacing $\sim 1/q$. In this limit the field $\phi$ varies
almost continuously and we can get rid of the lattice structure
and  use in the sequel the continuous notation.

 At the first order
one gets the  couplings
 \be
 {\cal L}_{\phi\psi\psi}=-\sum_{\vec
v}\frac{\pi\Delta}{R}\,\delta_R[h(\vec v\cdot\vec
n)]\frac{i\,\phi}{f}\,
 \epsilon_{ij}\epsilon^{\alpha\beta 3}\psi_{i,\alpha,\vec
v}\,C\,\psi_{j,\beta,-\vec v} -(L\to R)\ +\
h.c.\label{Trilineare}\ee
 \be
 {\cal L}_{\phi\phi\psi\psi}=\sum_{\vec
v}\frac{\pi\Delta}{R}\,\delta_R[h(\vec v\cdot\vec
n)]\frac{\phi^2}{2f^2}\ \epsilon_{ij}\epsilon^{\alpha\beta
3}\psi_{i,\alpha,\vec v}\,C\,\psi_{j,\beta,-\vec v}-(L\to R)\ +\
h.c.
 \label{quadrilineare}\ee
In the basis of the $\chi$ fields one has:
 \be {\mathcal L}_3\,+\,{\mathcal L}_4=
 \sum_{\vec v}\sum_{A=0}^3
\tilde\chi^{A\,\dag}\,   \left(
\begin{array}{cc}
 0 & g^\dag
\\
  g& 0\end{array}\right)
    \,\tilde\chi^B\ .\label{vertex}\ee
Here \be g=\left(\frac{\pi\Delta}{R}\,\delta_R[h(\vec v\cdot\vec
n)]\frac{i\,\phi}f\,-\frac{\pi\Delta}{R}\,\delta_R[h(\vec
v\cdot\vec n]\frac{\phi^2}{2f^2}\right)\left(\begin{array}{cccc}
   1& 0 & 0 & 0 \\
  0 & -1 &0 & 0 \\
  0 & 0&-1& 0 \\
  0 & 0 & 0 & -1
\end{array}\right)_{AB}\ee

To compute the effective lagrangian for the phonon field we use
the propagator given in Eq. (\ref{propagatore}) and the
interaction vertices in (\ref{vertex}). The result of the
calculation at the second order in the momentum expansion is
\footnote{One can easily control that the Goldstone theorem is
satisfied and the phonon is massless.} \be {\cal L}_{eff}(p)=i\,
 \frac{4\times 4\,\mu^2}{16\pi^3f^2} \sum_{\vec
v}\frac 1 2\left(\frac{\pi\Delta}{R}\,\delta_R[h(\vec v\cdot\vec
n_m)](i\,\phi)\right)^2\int d^2\ell\,\frac{2\Delta_{eff}^2\,
V\cdot p\,\tilde V\cdot p}{[D(\ell)]^3}\ . \ee One can handle the
fat delta according to the Fermi trick in the Golden Rule; in
expressions involving the gap parameters one makes in the
numerator the substitution $ \delta_R[h(x)]\to\delta[h(x)]$; one
fat delta is substituted by the Dirac delta while the other fat
delta gives $\displaystyle \frac{\pi\delta_R[h(x)]}
R\to\frac{\pi\delta_R(0)}R\to \,1 .$ We finally get the effective
lagrangian in the form
 \be{\cal L}_{eff}(p)\,=\,-\,\frac{\mu^2k_R}{2\pi^2f^2}
 \sum_{\vec v}
\delta\left\{\vec v\cdot\vec n-\frac{\delta \mu}{q}\right\}
V_\mu\tilde V_\nu p_\mu\phi p_\nu\phi\ . \ee Here $k_R$ is
kinematical factor of the order of 1 induced by the approximation
of the Riemann-Lebesgue lemma \cite{gattoloff}. The integration
over fermi velocities can be easily performed and  one obtains the
effective lagrangian in the form
 \be{\cal
L}(\phi,\partial_\mu\phi)=\frac{1}{2}\left( \dot\phi^2-
v_\parallel^2|\vec\nabla_\parallel\phi|^2-
v_\perp^2|\vec\nabla_\perp\phi|^2 \right)\ , \ee  if \be
 f^2= \frac {\mu^2k_R}{2\pi^2}
  \ .
\ee Here $\vec\nabla_\parallel\phi=\vec n(\vec
n\cdot\vec\nabla)\phi$ ,
$\vec\nabla_\perp\phi=\vec\nabla\phi-\vec\nabla_\parallel\phi$.
Moreover
 \be
v_\perp^2=\,\frac 1
 2\sin^2\theta_q\ ,\ee\be  v_\parallel^2=\,\cos^2\theta_q\ee and
 \be\displaystyle\cos^2\theta_q=\left(\frac{\delta
\mu}{q}\right)^2\ . \ee Therefore the dispersion law for the
phonon is anisotropic.

\section*{Acknowledgments}It is a pleasure to thank  R. Casalbuoni,
 R. Gatto and M.
Mannarelli for a very pleasant scientific collaboration on the
themes of these lectures and the organizers of the GISELDA
workshop at Frascati for an interesting and successful meeting.

\end{document}